\documentclass[aip,reprint,twocolumn,superscriptaddress,floatfix]{revtex4-1}
\usepackage{epsfig}
\usepackage{amsmath}
\usepackage{amssymb}
\usepackage{amsfonts}
\usepackage{mathptmx}
\usepackage{dcolumn}
\usepackage{eucal}
\usepackage{bm}
\usepackage{color}
\usepackage[colorlinks,linkcolor=blue,citecolor=blue]{hyperref}
\usepackage{epstopdf}
\usepackage{graphicx}
\usepackage{natbib}

\begin{document}
\title{Normal metal tunnel junction-based superconducting quantum interference proximity transistor: the N-SQUIPT}
\author{Sophie D'Ambrosio}
\email{sophie.dambrosio@nano.cnr.it}
\affiliation{NEST Istituto
Nanoscienze-CNR  and Scuola Normale Superiore, I-56127 Pisa, Italy}
\author{Martin Meissner}
\affiliation{NEST Istituto Nanoscienze-CNR  and Scuola Normale
Superiore, I-56127 Pisa, Italy}
\author{Christophe Blanc}
\affiliation{NEST Istituto Nanoscienze-CNR  and Scuola Normale
Superiore, I-56127 Pisa, Italy}
\author{Alberto Ronzani}
\affiliation{NEST Istituto Nanoscienze-CNR  and Scuola Normale
Superiore, I-56127 Pisa, Italy}
\author{Francesco Giazotto}
\email{francesco.giazotto@sns.it}
\affiliation{NEST Istituto Nanoscienze-CNR  and Scuola Normale
Superiore, I-56127 Pisa, Italy}

\begin{abstract}
We report the fabrication and characterization of an
alternative design for a superconducting quantum interference proximity
transistor (SQUIPT) based on a normal metal (N) probe. The absence of direct Josephson coupling between the
proximized metal nanowire and the N probe allows us to
observe  the full modulation of the wire density of states around
zero voltage and current \textit{via} the application of an external magnetic field. This results into a drastic suppression of power
dissipation which can be as low as a few $\sim 10^{-17}$ W. In this context the interferometer allows an improvement of up to
four orders of magnitude with respect to earlier SQUIPT designs, and makes it ideal for extra-low
power cryogenic applications. In addition, the N-SQUIPT has
been recently predicted to be the enabling candidate for the implementation of coherent caloritronic devices based on proximity effect.

\end{abstract}

\maketitle
\begin{figure}[t!]
\includegraphics[height=8.5cm, width=6.8cm]{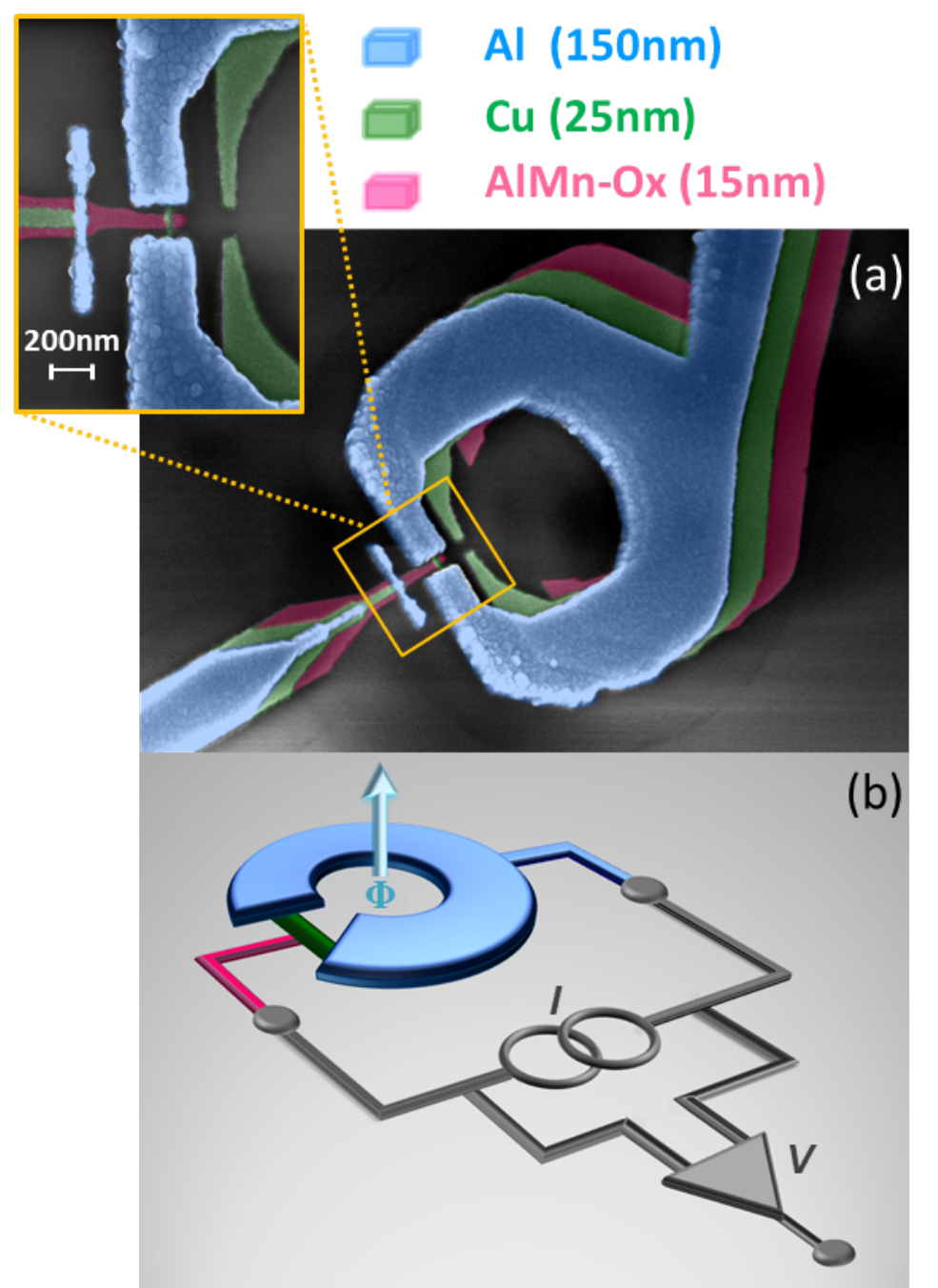}
\vspace{-0mm} \caption{(Color online) (a) Pseudo-color tilted scanning electron
micrograph (SEM) of a typical N-SQUIPT. The Cu normal metal wire is in clean metallic contact with the Al superconductive ring. An
Al$_{0.98}$Mn$_{0.02}$ normal metal tunnel probe is connected to
the middle of the Cu nanowire. The zoomed image on the left inset emphasises the core of the device close to the SNS
weak link. (b) Sketch of N-SQUIPT current bias measurement setup under fixed current-bias (\textit{\textsf{I}}). $\Phi$ symbolizes the externally applied magnetic flux piercing the loop whereas \textit{\textsf{V}} is the voltage drop. \label{fig1}} \vspace{-2mm}
\end{figure}

The superconducting quantum interference proximity transistor
(SQUIPT) is a magnetic-flux detector alternative to the widespread superconducting quantum interference device (SQUID)\cite{artalb1}.
Based on the proximity effect \cite{artalb2,cnrs2,cnrs23,prox1,prox2,prox4}, it is considered a promising
candidate for several advanced applications such as the next
generation of ultra-high sensitive and ultra-low power
magnetometers\cite{artalb9,artalb10,artalb11,alberto16}. The SQUIPT is a two-terminal device made of a normal
metal (N) nanowire embedded into a superconducting (S) ring, and coupled \emph{via} a tunnel barrier to a probing electrode (see
Fig.~\ref{fig1}). The N nanowire, in clean metallic contact
with the S ring, is proximized by the latter, and forms a SNS Josephson weak link. In this configuration, the
density of states (DOS) of the N wire is modulated  by the application of an external  magnetic flux $\Phi$ piercing  the loop, and enables the
transition of the wire from the N- to the S-like state\cite{artalb3,artalb4,artalb5,artalb6,artalb7,artalbPetro1,artalbPetro2,artalbPetro3,artalbZaikin1,artalbZaikin2,correctionALB3}.

Since its original introduction, the SQUIPT has been exclusively implemented
with a tunnel superconducting probe (S-SQUIPT) because of its sharper response and improved noise performance\cite{artalb1,artalb9,artalb10,artalb11,alberto16,rem2}.
Yet, it has been recently predicted that coherent thermal valves based on the proximity effect privilege SQUIPTs realized with a normal metal probe (N-SQUIPT), as the presence of the superconducting junction in the conventional S-SQUIPT design would severely limit the heat flow across the structure\cite{elia1}.
The N-SQUIPT appears therefore as a highly-promising candidate to implement future phase-coherent caloritronic devices such as
heat transistors, rectifiers, thermal splitters and phase-tunable electron
coolers \cite{RMP2006}.
Besides the foreseen advantages in coherent caloritronics, the N-SQUIPT shows
attractive performance for more conventional electronic applications due to the lack of Josephson coupling. The N probe offers the possibility to operate the device around zero bias therefore allowing to reach extra-low power dissipation, down to a few tens of aW, which lowers by up to four orders of
magnitude the previous achieved dissipation values\cite{artalb10,artalb11,alberto16}.

Here we report the fabrication and the magneto-electric characterization of Al/Cu-based N-SQUIPTs. After describing the different steps required to realize this device, we will show the full electric behavior as a function of bath temperature. Our results are reproduced with the Usadel equations describing the proximity effect in the N wire. Moreover, the N-SQUIPTs are characterized by a maximum flux-to-voltage transfer function of $\sim 0.45$ mV/$\Phi_{0}$ and maximum flux-to-current transfer function of $\sim 12$ nA/$\Phi_{0}$.

Figure~\ref{fig1} (a) shows a scanning electron microscopy (SEM) image
of a typical N-SQUIPT device. The samples have been
fabricated through electron-beam lithography (EBL) and three-angle shadow-mask evaporation performed on an
oxidized Si wafer covered with a suspended bilayer resist mask made
of 900-nm-thick copolymer layer and 300-nm-thick polymethyl
methacrylate (PMMA) layer spun on top of it. The EBL step is followed by development in
methyl isobutyl ketone:isopropanol (MIBK:IPA) 1:3 solution for 1 min,
rinsing in IPA and then drying. The different metal layers
are deposited by ultra-high vacuum electron-beam evaporation at different angles. At first 15 nm of Al$_{0.98}$Mn$_{0.02}$ are deposited at
40$^{\circ}$ and oxidized for 5 min with an oxygen pressure of 37
mTorr to realize the probe electrode, then 25 nm of Cu
are evaporated at 20$^{\circ}$ to form the proximized N wire, and finally a 150-nm-thick Al superconductor ring is deposited at zero angle. The oxidation of the N probe is crucial to make functional our device. It insures a well-defined voltage bias allowing the measurement of the DOS, and avoids as well any weakening effect due to the inverse proximity effect. Furthermore concerning the caloritronic applications, the control of the resistivity through the tunnel barrier allows us to limit the power dissipation of our system. To achieve full phase polarizability the SQUIPT
requires to have an S ring with a thickness much larger
than the wire in order to reach the condition
$L^{R}\ll L^{WL}$, where \emph{L}$^{R(WL)}$ denotes the inductance of the ring(weak link)\cite{artalb10,artalb11,alberto16}. The value of the ring inductance has been estimated with the finite-elements software FastHenry\cite{correctionALB1} to be less than 5 pH.

We have measured our N-SQUIPTs in a He$^{3}$-He$^{4}$
dilution refrigerator at different temperatures ranging from 25 mK to 1.2 K
using room temperature preamplifiers. The main characteristics of our devices, summarized
in Tab.~\ref{tab1}, demonstrate
the good level of reproducibility achieved with the fabrication process described. In the following we report the measurements obtained for sample A, the other samples showing similar results.
\begin{table}[t!]
\begin{center}
 \begin{tabular}{l c c c c c c}
 \hline
 \hline
      & $l$ & $d$ & $w$ & $R_{T}$ & $|\textit{\textsf{dI/d}}{\Phi}|_{\textsf{Max}}$ & $|\textit{\textsf{dV/d}}\Phi|_{\textsf{Max}}$ \\
 Sample & (nm) & (nm) & (nm) & (k$\Omega$) & (nA/$\Phi_{0}$) & (mV/$\Phi_{0}$)\\
 \hline
 A & 160 & 60 & 120 & 33 & 12 & 0.45\\

 B & 160 & 55 & 120 & 36 & 11.5 & 0.41 \\

 C & 150 & 60 & 90 & 40 & 12 & 0.46 \\
 \hline
 \hline
\end{tabular}
\end{center}
\caption{\label{tab1} Parameters of three different N-SQUIPT samples measured at $\textit{\textsf{T}}_{\textit{\textsf{bath}}}=25$ mK.  The symbols $l$ and $d$ are
used to denote the length and the width of the N nanowire, respectively,
whereas $w$ indicates the width of N probe. $R_{T}$ is the normal-state tunnel junction resistance of the N probe. $|\textit{\textsf{dI/d}}\Phi|_{\textsf{Max}}$ and $|\textit{\textsf{dV/d}}\Phi|_{\textsf{Max}}$ are the maximum absolute value of the flux-to-current and
flux-to-voltage transfer functions, respectively.}
\end{table}
From the length $l$ of the N wire, we deduce the Thouless energy
$E_{Th}=\hbar D /l^{2}\simeq0.8\;\Delta_{0}$, by using $D\simeq60$
cm$^{2}$s$^{-1}$ for the diffusion coefficient in the Cu wire, and
$\Delta_{0}\simeq190$ $\mu$eV as the zero-temperature energy gap in the Al loop which are estimated from previous works\cite{alberto16}.
The above given value for the ratio $E_{Th}/\Delta_{0}$ sets the frame of the \textit{intermediate-length} junction regime of the SNS weak link.

\begin{figure}[t!]
\includegraphics[height=17cm, width=9cm]{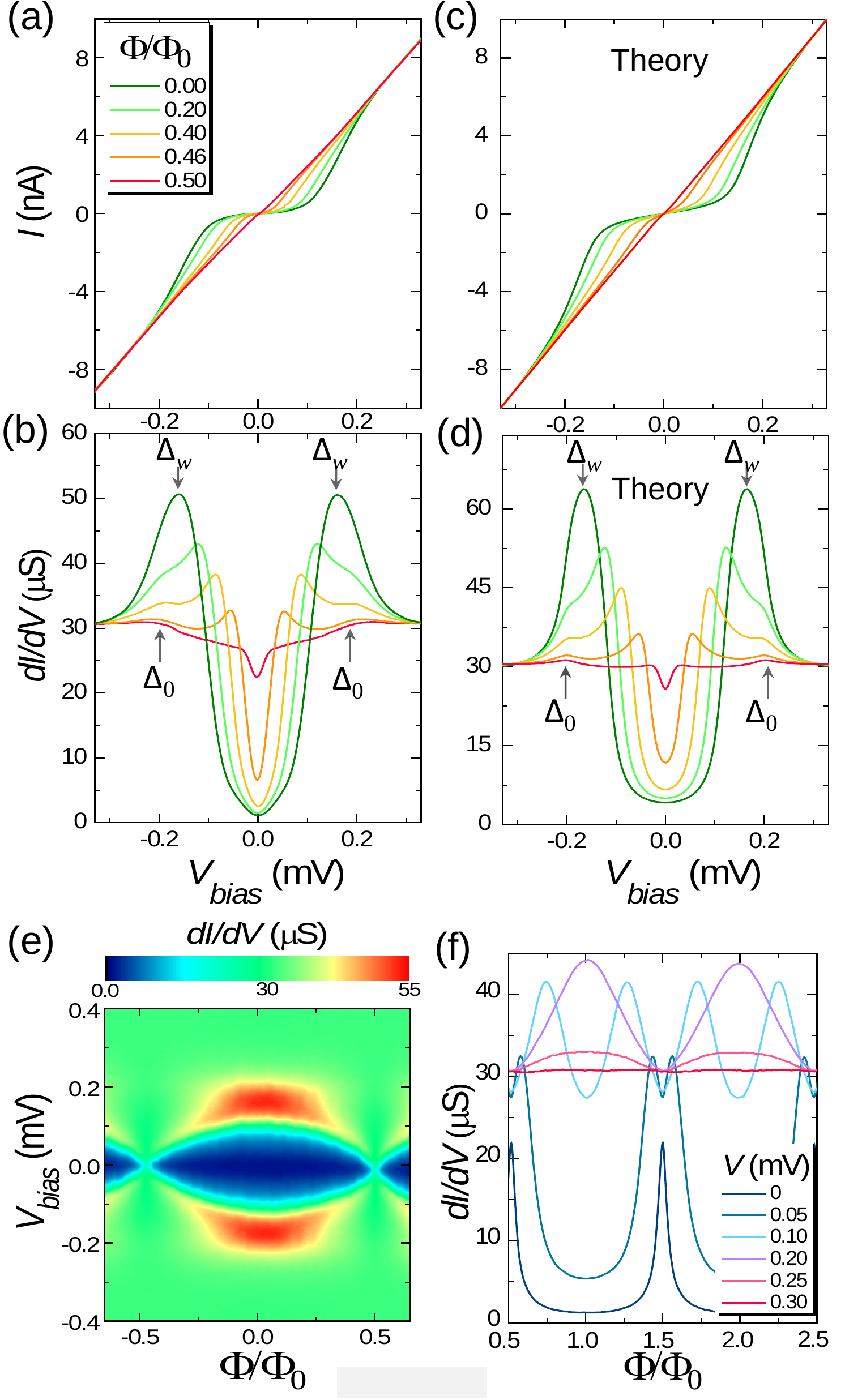}
\vspace{-0mm} \caption{(Color online) (a) Current versus voltage \textit{\textsf{I}}(\textit{\textsf{V}}$_{\textit{\textsf{bias}}}$)
measured at different magnetic flux $\Phi$ and \textit{\textsf{T}}$_{\textit{\textsf{bath}}}$=25
mK. (b) Differential conductance versus voltage\textit{\textsf{ dI/dV}}(\textit{\textsf{V}}$_{\textit{\textsf{bias}}}$)
measured for $\Phi$ and
\textit{\textsf{T}}$_{\textit{\textsf{bath}}}$ as in (a). (c) and (d) Theoretical calculations of
the curves shown in panels (a) and (b), respectively. $\Delta_{0}$ is the zero-temperature energy gap in the Al loop and $\Delta_{w}$ the maximum induced minigap in the Cu nanowire at $\Phi=0$. (e) Color
plot of the differential conductance versus voltage and magnetic flux
\textit{\textsf{dI/dV}}(\textit{\textsf{V}}$_{\textit{\textsf{bias}}}$,$\Phi$) measured at \textit{\textsf{T}}$_{\textit{\textsf{bath}}}$=25 mK. (f) Differential conductance versus magnetic flux \textit{\textsf{dI/dV}}($\Phi$) measured at various
voltages and \textit{\textsf{T}}$_{\textit{\textsf{bath}}}$=25 mK.\label{fig2}}
\vspace{-2mm}
\end{figure}
Figures~\ref{fig2}(a) and (b) show the current
\textit{\textsf{I}}(\textit{\textsf{V}}$_{\textit{\textsf{bias}}}$)
and the differential conductance
\textit{\textsf{dI/dV}}(\textit{\textsf{V}}$_{\textit{\textsf{bias}}}$)
measured for different magnetic flux $\Phi$ applied orthogonally to the ring at
$\textit{\textsf{T}}_{\textit{\textsf{bath}}}=25$ mK [see Fig.~\ref{fig1}(b)]. Data show
evidence of the full wire DOS modulation, the N wire going from
the S-like state with a maximum induced minigap $\Delta_{w}\simeq
160$ $\mu$eV at $\Phi=0$, to the N-like state $\Delta_{w}\simeq 0$
at $\Phi/\Phi_{0}=0.5$, where $\Phi_{0}= 2.067\times10^{-15}$ Wb is
the flux quantum. From the curves of the differential conductance
displayed in Fig.~\ref{fig2}(b), we can estimate the Al ring
gap $\Delta_{0}\simeq190$ $\mu$eV which appears as a peak at higher voltage for $\Phi\neq0$ and the
maximum value of the Cu minigap $\Delta_{w}\simeq160$ $\mu$eV corresponding to the position of the peaks at $\Phi=0$, as shown by the arrows in Fig.~\ref{fig2}(b). Both
values are consistent with the prediction obtained from the solution of the
quasiclassical equations as explained below.

The overall theoretical comparison to the experimental data of
Figs.~\ref{fig2}(a) and (b) is presented in Figs.~\ref{fig2}(c) and
(d), respectively, and demonstrates a fairly good qualitative
agreement with the experiment. These calculations have
been obtained from the solution of the Usadel equations
\cite{usadel} which describes the proximity effect in the diffusive N
wire, and can be written as
\begin{equation}
\label{eq2}
\hbar D\partial_{x}^{2}{\gamma_{1,2}}-\hbar D\dfrac{2\gamma_{2,1}}{1+\gamma_{1}\,\gamma_{2}}(\partial_{x}\,{\gamma_{1,2}})^{2}
+2i(E+i\Gamma)\gamma_{1,2}=0,
\end{equation}
where the first equation corresponds to the first index of $\gamma$
and the second equation to the second
index\cite{bergeret,franzusadel}. The functions $\gamma_{1,2}$
determined with Riccati parametrization\cite{riccati1,riccati2}
are used to describe the retarded Green function
$\mathcal{G}(x,E,\Phi,\textit{\textsf{T}})=(1-\gamma_{1}\gamma_{2})/(1+\gamma_{1}\gamma_{2})$,
and are dependent of the position $x$, the energy $E$, the
flux $\Phi$, and the temperature $\textit{\textsf{T}}$ as well. The
DOSs in the normal proximized region can be expressed by
$\mathcal{N}_{N}(x,E,\Phi,\textit{\textsf{T}})=\textrm{Re}[\mathcal{G}(x,E,\Phi,\textit{\textsf{T}})]$.
The value $\Gamma=0.065\;\Delta_{0}$ is taken as input parameter,
and describes the inelastic scattering present in the N
region. Here the Nazarov boundary
conditions\cite{bergeret38,bergeret39} are used in order to include in the simulation details such as the quality of the normal
metal-superconductor interfaces in the SNS weak
link\cite{boundaryexplain}. The current flowing through the N probe can be written as
\begin{equation}
\label{eq1}
\begin{split}
\textit{\textsf{I}}\;(\Phi,\textit{\textsf{V}},\textit{\textsf{T}})=\frac{1}{eR_Tw} &\;\int_{0}^{w}dx\;\,\int_{-\infty}^{\infty}\;\, dE\;\, \mathcal{N}_{N}(x,E,\Phi,\textit{\textsf{T}})\;\;\;\\
 &\;\;\;\;\;\times \left[\mathcal{F}_{0}(E-e\textit{\textsf{V}},\textit{\textsf{T}})-\mathcal{F}_{0}(E,\textit{\textsf{T}})\right],
\end{split}
\end{equation}
where $e$ is the electron charge, $\mathcal{F}_{0}$ is the Fermi distribution function and
$\mathcal{N}_{N}(x,E,\Phi,\textit{\textsf{T}})$ is calculated by solving Eqs.~\ref{eq2}.
The \textit{\textsf{I}}(\textit{\textsf{V}}) curves shown in
Fig.~\ref{fig2}(c) directly correspond to the calculation of
Eq.~\ref{eq1} with $\textit{\textsf{T}}_{\textit{\textsf{bath}}}=25$
mK and $\Phi$ values as in Fig.~\ref{fig2}(a). The
differential conductances traces displayed Fig.~\ref{fig2}(d) are the derivatives of the curves shown in Fig.~\ref{fig2}(c). From Fig.~\ref{fig2}(d), we confirm the value $\Delta_{0}\simeq190$ $\mu$eV of the S ring which corresponds to a critical temperature $\textit{\textsf{T}}_{C}=\Delta_{0}/(1.764 k_{B})\simeq1.2$ K. We extract as well the effective N wire minigap $\Delta_{w}\simeq160$ $\mu$eV at $\Phi=0$.

The color plot of $\textit{\textsf{dI/dV}}(\textit{\textsf{V}}_{\textit{\textsf{bias}}},\Phi)$ shown in Fig.~\ref{fig2}(e) gives a direct observation of the DOS modulation which goes from a fully developed minigap at $\Phi=0$, to an almost closed minigap at $|\Phi/\Phi_{0}|=0.5$ with a behavior which is $\Phi_{0}$-periodic in the magnetic flux\cite{artalb2,tinkham}.

Figure~\ref{fig2}(f) shows the differential conductance $\textit{\textsf{dI/dV}}(\Phi)$ at different values $\textit{\textsf{V}}_{\textit{\textsf{bias}}}$, and provides the experimental evidence that an appreciable magnetic flux response in the differential conductance values is obtained with an extra-low dissipation measurement setup. Indeed, the observed modulations have been measured with a lock-in amplifier with an input amplitude modulation $\textit{\textsf{V}}_{AC}\simeq10^{-6}$ V. At $\textit{\textsf{V}}=0$ the typical output current level is $\textit{\textsf{I}}_{AC}\simeq 10^{-11}$ A with an average total power dissipation for the N-SQUIPT around $10^{-17}$ W. The typical value of the dynamic resistance (on the order of 30 k$\Omega$) sets the limit of the reachable bandwidth depending on the shunt capacitance typically due to line filtering\cite{correctionALBfinal}.

\begin{figure}[t!]
\includegraphics[height=14.5cm, width=9cm]{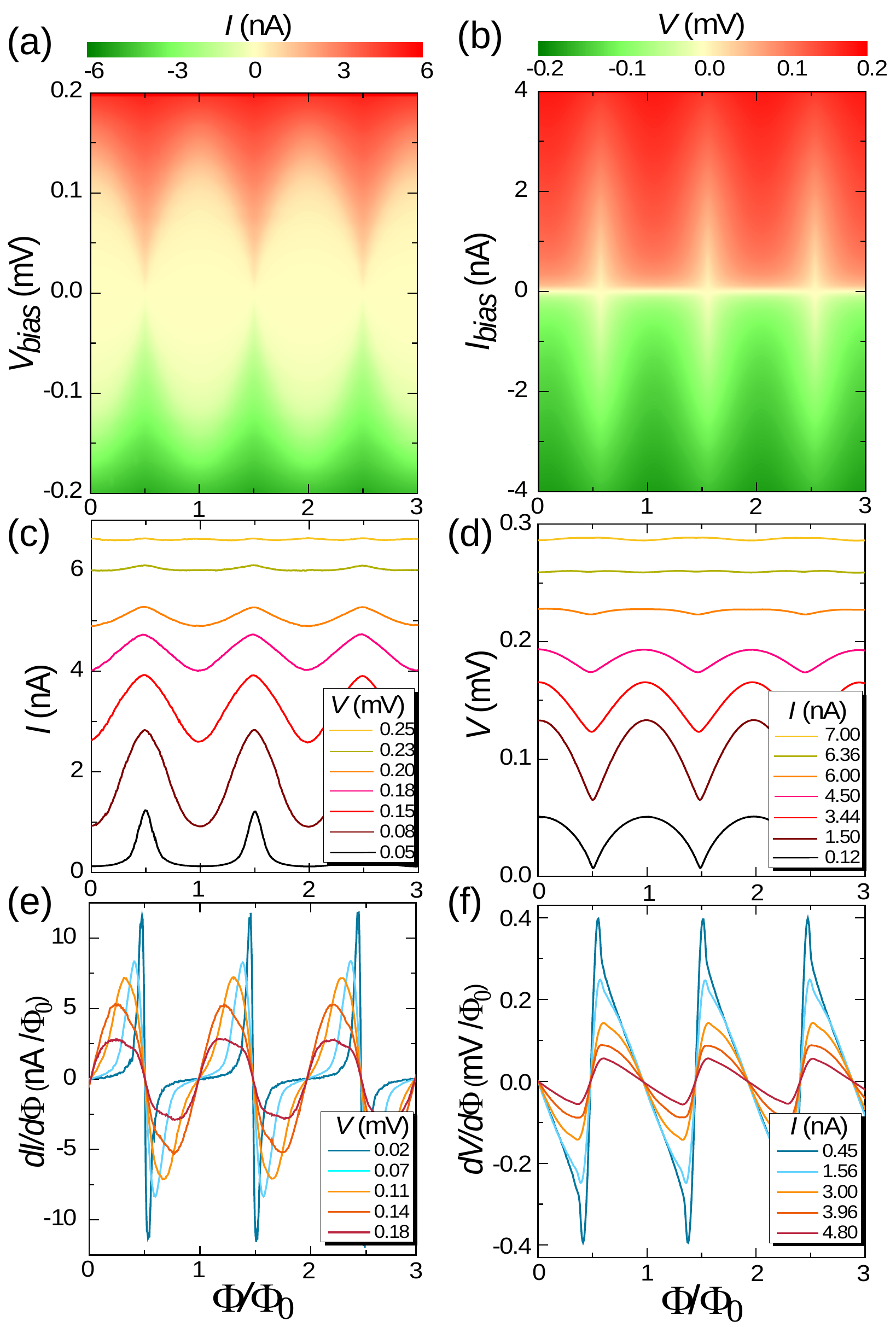}
\vspace{-8mm}
\caption{(Color online) (a) and (b) Color plots which present the flux-modulated current as a function
of voltage $\textit{\textsf{I}}(\textit{\textsf{V}}_{\textit{\textsf{bias}}},\Phi)$
and the flux-modulated voltage as a function of current $\textit{\textsf{V}}(\textit{\textsf{I}}_{\textit{\textsf{bias}}},\Phi)$
at \textit{\textsf{T}}$_{\textit{\textsf{bath}}}$=25 mK, respectively. (c) Curves of flux-modulated current at different voltage values selected from (a). (d) Curves of flux-modulated voltage at different current values selected from (b).
(e) and (f) Flux-to-current ($\textit{\textsf{dI/d}}{\Phi}$),
and flux-to-voltage ($\textit{\textsf{dV/d}}{\Phi}$) transfer function curves obtained from the derivative of the \textit{\textsf{I}}($\Phi$) and \textit{\textsf{V}}($\Phi$) measurements, respectively.
\label{fig3}
}
\vspace{-2mm}
\end{figure}
Figure~\ref{fig3} completes the electrical characterization of our device. In particular, Figs.~\ref{fig3}(a) and (b) show the color plots of $\textit{\textsf{I}}(\textit{\textsf{V}}_{\textit{\textsf{bias}}},\Phi)$ and $\textit{\textsf{V}}(\textit{\textsf{I}}_{\textit{\textsf{bias}}},\Phi)$, respectively. Figures~\ref{fig3}(c) and (d) display a cross section of the above color plots for some selected values of $\textit{\textsf{V}}_{\textit{\textsf{bias}}}$ and $\textit{\textsf{I}}_{\textit{\textsf{bias}}}$. A careful inspection of Figs.~\ref{fig3}(a) and (b) indicates that the measured current and voltage modulations reach peak-to-peak amplitudes as large as $\delta \textit{\textsf{I}}=2$ nA and $\delta \textit{\textsf{V}}\simeq90$ $\mu$V, respectively. We note that the lower value for $\delta \textit{\textsf{V}}$ in comparison to the full minigap amplitude $\Delta_{w}\simeq160$ $\mu$eV can be reproduced in our simulation by including terms related to a finite inelastic scattering in the N wire and to a non ideal Al/Cu interface transmissivity\cite{boundaryexplain} [see Figs.~\ref{fig2}(c) and (d)].

The SQUIPT behaves as a flux-to-current or a flux-to-voltage transformer whose response efficiency can be quantified by the maximum absolute value of its flux-to-current ($|\textit{\textsf{dI/d}}\Phi|_{\textsf{Max}}$) or flux-to-voltage ($|\textit{\textsf{dV/d}}\Phi|_{\textsf{Max}}$) transfer functions, respectively\cite{artalb1,artalb9,artalb10,artalb11,alberto16}. In our case, this information is shown in Figs.~\ref{fig3}(e) and (f), respectively. Their values reach $|\textit{\textsf{dV/d}}\Phi|_{\textsf{Max}}\simeq0.45$ mV/$\Phi_{0}$ and $|\textit{\textsf{dI/d}}{\Phi}|_{\textsf{Max}}\simeq12$ nA/$\Phi_{0}$ at 25 mK, respectively (see Tab.~\ref{tab1}). Although higher values have been reported in S-SQUIPTs \cite{artalb11,alberto16}, the N-SQUIPTs still exhibit performance on par with conventional state-of-art SQUID sensors\cite{handbookSQUID}.

We now discuss the noise-equivalent-flux (NEF) or flux resolution
($\Phi_{NS}$) of the N-SQUIPT. The intermediate value of the tunnel
junction impedance allows the devices to be operated either with
voltage amplification under DC current bias or with current
amplification under DC voltage bias. In the former configuration a
maximum voltage responsivity $\left| \mathrm{\textit{\textsf{dV/d}}\Phi}
\right|_{\textsf{Max}} \simeq 0.45\, \mathrm{mV/\Phi_0}$ has been recorded with
$\textit{\textsf{I}}_{\textit{\textsf{bias}}} \simeq 400\, \mathrm{pA}$, corresponding to $\Phi_{NS} =
\sqrt{S_\textit{\textsf{V}}} / \left| \mathrm{\textit{\textsf{dV/d}}\Phi} \right|_{\textit{\textsf{Max}}} \simeq 3.4\, \mu
\Phi_0 / \sqrt{\mathrm{Hz}}$, where $\sqrt{S_\textit{\textsf{V}}}$ is the input-referred
noise power spectral density of the preamplifier used in this setup\cite{albfootnote}.
Improved performance can be obtained by exploiting the low
input-referred noise level granted by a transimpedance
current preamplifier\cite{fn1} combined
with the significant current responsivity $\left| \mathrm{\textit{\textsf{dI/d}}
\Phi} \right|_{\textsf{Max}} \simeq 12\, \mathrm{nA/\Phi_0}$ achieved at $\textit{\textsf{V}}_{\textit{\textsf{bias}}}
\simeq20\, \mathrm{\mu V}$. In this configuration the achievable magnetic
flux resolution is expected to be limited by the shot noise of the
tunnel junction, reaching values as low as $\Phi_{NS} = \sqrt{2e\textit{\textsf{I}}} /
\left| \mathrm{\textit{\textsf{dI/d}}\Phi}\right|_{\textsf{Max}} \simeq 1.5\, \mu \Phi_0 /
\sqrt{\mathrm{Hz}}$, where $\textit{\textsf{I}} \simeq 1\,
\mathrm{nA}$. The magnitude of the dissipation induced in the DC
readout is of the order of tens of fW. We notice that the contribution due to the presence of a finite ring inductance in the noise performance is negligible\cite{correctionALB2}.

As already noted, the N-SQUIPT can also be operated at zero DC bias, where its response can be
linearized. In this configuration the operating power can be brought
down to the aW range by applying a minute AC excitation, with
the same technique which is used in resistance bridges adapted to
cryogenic thermometry. The maximal zero-bias conductance responsivity
estimated from the data shown in Fig.~\ref{fig2}(f) is $\left| \mathrm{\textit{\textsf{dG}}_0/\textit{\textsf{d}}
\Phi} \right|_{\textit{\textsf{Max}}} \simeq 450\, \mathrm{\mu S/\Phi_0}$, leading to a
magnetic flux resolution $\Phi_{NS} = \sqrt{S_\textit{\textsf{G}}} / \left| \mathrm{\textit{\textsf{dG}}_0/\textit{\textsf{d}}\Phi} \right|_{\textit{\textsf{Max}}} \simeq 300\, \mu \Phi_0 / \sqrt{\mathrm{Hz}}$,
where $\sqrt{S_\textit{\textsf{G}}} \simeq 140 \, \mathrm{nS/\sqrt{Hz}}$
 is the
noise-equivalent power spectral density of the lock-in amplification
setup. This value of the flux resolution has been obtained with a
1-$\mu V$ AC voltage excitation, corresponding to $\sim17$ aW of applied
power.
\begin{figure}[t!]
\includegraphics[height=10.5cm, width=9cm]{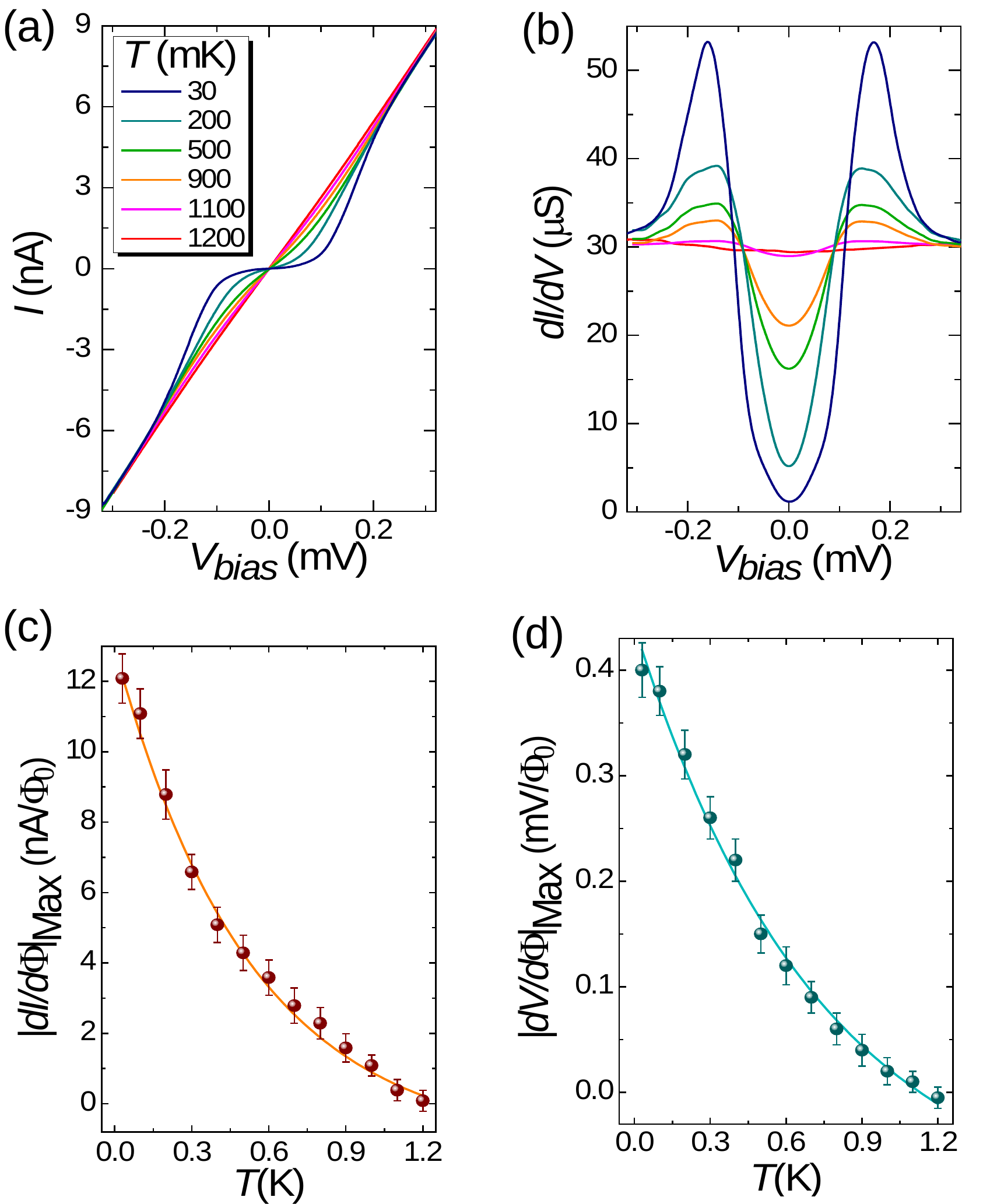}
\vspace{-2mm}
\caption{(Color online) (a) Current versus voltage \textit{\textsf{I}}(\textit{\textsf{V}}$_{\textit{\textsf{bias}}}$)
measured for various bath temperatures $\textit{\textsf{T}}_{\textit{\textsf{bath}}}$ at $\Phi=0$. (b) Differential conductance \textit{\textsf{dI/dV}}(\textit{\textsf{V}}$_{\textit{\textsf{bias}}}$) for the same temperatures as in (a). (c) and (d) Temperature dependence of the maximum flux-to-current and flux-to-voltage transfer functions, respectively.
The error bars shown in (c) and (d) represent the standard deviation of maximum values estimated from $\textit{\textsf{I}}(\Phi)$ and $\textit{V}(\Phi)$ data over several flux periods. The lines connecting the experimental data in (c) and (d) are an interpolation to guide the eye. \label{fig4}
}
\vspace{-0mm}
\end{figure}
The impact of temperature $\textit{\textsf{T}}$
is displayed in Fig.~\ref{fig4}. In particular, Fig.~\ref{fig4}(a)
shows the evolution of
$\textit{\textsf{I}}(\textit{V}_{\textit{\textsf{bias}}})$
at different temperatures when the minigap in the N wire DOS is fully
developed (i.e. at $\Phi=0$). As expected, when \textit{\textsf{T}}
increases, the proximity effect in the SNS junction is progressively
weakened and completely disappears at 1.2 K which
corresponds to the critical temperature of
the Al ring. Figure~\ref{fig4}(b) shows the corresponding
differential conductance and confirms the suppression of the minigap
at $\textit{\textsf{T}}_{C}$.
The increase in temperature leads to a reduction of $\textit{V}(\Phi)$ and
$\textit{\textsf{I}}(\Phi)$ which, as a consequence, suppresses the amplitude of
the flux-to-voltage and flux-to-current transfer functions, as shown in
Figs.~\ref{fig4}(c) and (d), respectively. We emphasize that the N-SQUIPT shows appreciable values for both the maximum flux-to-voltage and flux-to-current transfer functions even at somewhat high bath temperatures. In particular, at 1 K our interferometers still exhibit $|\textit{\textsf{dI/d}}{\Phi}|_{\textmd{\textsf{Max}}}\simeq1$ nA/$\Phi_{0}$ and $|\textit{\textsf{dV/d}}{\Phi}|_{\textit{\textsf{Max}}}\simeq30$ $\mu$V/$\Phi_{0}$.

In summary, we have performed the fabrication and the
magneto-electrical characterization of Al/Cu-based N-SQUIPTs. The design choice of the N probe characterized by the absence of the Josephson coupling with the proximized weak link shows several advantages: i) the transition between a fully linear to a highly nonlinear characteristic (unlike the S-SQUIPT in which only nonlinear behavior is possible) suggests the adoption of the N-SQUIPT as a fully metallic, highly-efficient, tunable electrical diode for operation at temperature below 1K. ii) The typical operating power, fully modulated from fW to aW levels, gives the opportunity of using the N-SQUIPT as a magnetometer for condensed matter systems \cite{bauer} characterized by low-energy excitations, which are vulnerable to disruption by measurement back-action. iii) The choice of a normal metal probe has been shown to improve drastically the transport properties of a heat nanovalve \cite{elia1} due to the lack of a superconducting gap in the probe itself. More generally, a normal density of states provides a natural opportunity of realizing a thermal reservoir in which the electron temperature can be tuned and probed to its fullest extent. These properties make the N-SQUIPT a privileged building block for the implementation of coherent caloritronic devices based on proximity effect.

The authors thanks C. Altimiras, F. S. Bergeret and E. Strambini for fruitful comments.
The European Research Council under the European Union's Seventh Framework Program (FP7/2007-2013)/ERC Grant agreement No. 615187-COMANCHE and MIUR-FIRB2013 -- Project Coca (Grant No.~RBFR1379UX) are acknowledge for partial financial support.


\begin{thebibliography}{99}
\bibitem{boundaryexplain}
The Nazarov boundaries conditions introduce two additional
parameters $r$ and $\tau$ which correspond to the transparency and
the transmission coefficients of the SN interfaces, respectively. In
the convention used here an ideal interface is characterized by
$r=0$ and $\tau=1$. In our case, the calculations shown
Figs.~\ref{fig2}(c) and (d) are obtained with $r=0.33$ and
$\tau=0.5\,$.
\bibitem{correctionALBfinal}
Here the bandwidth is limited to 60 Hz due to the presence of  a 50 nF capacitance in the RF filters. This limit is not intrinsic to the device and can be lifted with cryogenic wiring and filtering optimized for this application.
\end{thebibliography}
\end{document}